\documentclass[conference, compsoc]{IEEEtran}
\usepackage{graphicx}
\newcommand{\be}{\begin{equation}}
\newcommand{\ee}{\end{equation}}
\newcommand{\bra}{\langle}
\newcommand{\ket}{\rangle}
\newcommand{\bea}{\begin{eqnarray}}
\newcommand{\eea}{\end{eqnarray}}
\newcommand{\dis}{\displaystyle}

\ifCLASSINFOpdf
\else
\fi
\begin{document}

\title{Bayesian Inference on QGARCH Model Using the Adaptive Construction Scheme}

%
%\author{Tetsuya Takaishi}%

% the affiliations are given next
%\institute{Springer-Verlag, Computer Science Editorial,\\
%Tiergartenstr. 17, 69121 Heidelberg, Germany\\
%\institute{Hiroshima University of Economics,\\
%731-0192  Hiroshima, Japan\\
%\mailsa
%%\\
%%\url{http://www.springer.com/series/7911}
%}


\author{\IEEEauthorblockN{Tetsuya Takaishi}
\IEEEauthorblockA{Hiroshima University of Economics\\
Hiroshima, Japan\\
Email: takaishi@hiroshima-u.ac.jp}
}


\maketitle

\begin{abstract}
We study the performance of the adaptive construction scheme 
for a Bayesian inference on the Quadratic GARCH model which
introduces the asymmetry in time series dynamics. 
In the adaptive construction scheme 
a proposal density in the Metropolis-Hastings algorithm
is constructed 
adaptively by changing the parameters of the density
to fit the posterior density. 
Using artificial QGARCH data 
we infer the QGARCH parameters by applying the adaptive construction scheme
to the Bayesian inference of QGARCH model.
We find that the adaptive construction scheme 
samples QGARCH parameters effectively, i.e. 
correlations between the sampled data are very small. 
We conclude that the adaptive construction scheme
is an efficient method to the Bayesian estimation of the QGARCH model. 
\end{abstract}

\section{Introduction}
A notable feature of financial time series is that
volatility of asset returns varies in time 
and high (low) volatility persists, which is called volatility clustering. 
Moreover the return distributions show fat-tailed distributions. 
These empirical properties are called stylized facts.
There also exist further stylized facts seen in financial markets\cite{CONT}.

A primary importance in empirical finance is 
to make models which mimic the properties of the volatility and
then to forecast future volatility.   
The most successful model is the Generalized Autoregressive Conditional Heteroscedasticity (GARCH)
model by Engle\cite{ARCH} and Bollerslev\cite{GARCH}, 
which can capture the property of volatility clustering and show the fat-tailed distribution.

In the original GARCH model, the process generates symmetric time series.
For stock markets, however, stock returns may show significant negative skewness in volatility dynamics.  
In order to incorporate the skewness into the model,
some modified models have been proposed\cite{EGARCH,GJR,QGARCH1,QGARCH2}.
Among them we focus on the Quadratic GARCH (QGARCH) model\cite{QGARCH1,QGARCH2} which includes an additional term 
that can capture the property of the skewness.

A preferred algorithm to infer GARCH model parameters is the Maximum Likelihood (ML) method
which estimates the parameters by maximaizing the corresponding likelihood function of the GARCH model.
In this algorithm, however, there is a practical difficulty in the maximization procedure when the output results are sensitive to
starting values.

By the recent computer development the Bayesian inference implemeneted by Markov Chain Monte Carlo (MCMC) methods,
which is an alternative approach to estimate GARCH parameters,
has become popular.
There exists a variety of methods proposed to implement the MCMC scheme\cite{Bauwens}-\cite{HMC}.
In a recent survey\cite{ASAI} it is shown that Acceptance-Rejection/Metropolis-Hastings  (AR/MH) algorithm
works better than other algorithms.
In the AR/MH algorithm the proposal density is assumed to be a multivariate Student's t-distribution and
the parameters to specify the distribution are estimated by the ML technique.
Recently an alternative method to estimate those parameters without relying on the ML technique was proposed
\cite{ACS}.
In this method the parameters are determined by using the data generated by an MCMC method and
updated adaptively during the MCMC simulation.
We call this method "adaptive construction scheme".

The adaptive construction scheme was tested for artificial GARCH data and
it is shown that the adaptive construction scheme can significantly reduce
the correlation between sampled data\cite{ACS}.
In this study we apply the adaptive construction scheme to
the QGARCH model. The orignal GARCH model has 3 model parameters.
On other hand the QGARCH model has 4 model parameters.
We study the efficiency of the adaptive construction scheme 
applied to the QGARCH model and examine whether 
the efficiency is still high enough for such a model with many parameters.

\section{QGARCH Model}

The QGARCH model\cite{QGARCH1,QGARCH2} which we employ here is written as
\be
y_t=\sigma_t \epsilon_t,
\ee
\be
\sigma_t^2  = \omega + \gamma y_{t-1}+ \alpha y_{t-1}^2 + \beta \sigma_{t-1}^2,
\ee
where 
$\epsilon_t$ is an independent normal error $\sim N(0,1)$ 
and $y_t$ are observations.
%$\omega>0$, $\alpha_i>0$ and $\beta_i>0$ to ensure a positive volatility.
%Furthermore the stationary condition $\sum_{i=1}^{p}\alpha_i + \sum_{i=1}^{q}\beta_i <1$ is also required.
Here $\alpha,\beta, \gamma$ and $\omega$ are the parameters to be estimated in the Bayesian inference.
The QGARCH process differs from the GARCH one by the term $\gamma y_{t-1}$ which introduces 
asymmetry.  

\section{Bayesian inference}
From the Bayes' theorem  
the posterior density $\pi(\theta|y)$  
with $n$ observations $y=(y_1,y_2,\dots,y_n)$ is given by
\be
\pi(\theta|y)\propto L(y|\theta) \pi(\theta),
\ee
where $L(y|\theta)$ is the likelihood function.
$\pi(\theta)$ is the prior density for $\theta$.
The functional form of $\pi(\theta)$ is not known a priori.  
Here we assume that the prior density $\pi(\theta)$ is constant.

For the QGARCH model  the likelihood function is given by
\be
L(y|\theta)=\Pi_{i=1}^{n} \frac1{\sqrt{2\pi\sigma_t^2}}\exp\left.(-\frac{y_t^2}{\sigma_t^2}\right.),
\ee
where $\theta=(\omega,\alpha,\beta,\gamma)$ stands for the QGARCH parameters.

Using $\pi(\theta|y)$ the QGARCH parameters are inferred as the expectation values given by
\be
\bra {\bf \theta} \ket = \frac1{Z}\int {\bf \theta} \pi(\theta|y) d\theta,
\label{eq:int}
\ee
where $Z=\int \pi(\theta|y) d\theta$ is a normalization constant irrelevant
to MCMC estimations.

%\subsection{Markov Chain Monte Carlo}
The MCMC technique gives a method to estimate eq.(\ref{eq:int}) numerically.
The basic procedure of the MCMC method is as follows.
First we sample $\theta$ drawn from a probability distribution
$\pi(\theta|y)$. 
%Since usually the probability distribution is not a simple form like a Gaussian distribution,  
Sampling is done by a technique which produces a Markov chain.  
After sampling  some data, 
we evaluate the expectation value as an average value over the sampled data $\theta^{(i)}$, 
\be
\bra {\bf \theta} \ket = \lim_{k \rightarrow \infty} \frac1k\sum_{i=1}^k \theta^{(i)},
\ee
where 
$k$ is the number of the sampled data.
The statistical error for $k$ independent data 
is proportional to $\frac1{\sqrt{k}}$.
In general, however, the data generated by the MCMC method are
correlated. As a result the statistical error will be proportional to $\sqrt{\frac{2\tau}{k}}$ 
where $\tau$ is the autocorrelation time between the sampled data.
The autocorrelation time depends on the MCMC method we employ.
Thus it is desirable to choose a MCMC method which can generate data with a small $\tau$.

\subsection{Metropolis-Hastings algorithm}

The MH algorithm\cite{MH} is a generalized version of the Metropolis algorithm\cite{METRO}. 
Let us consider to generate data $x$ from a probability distribution $P(x)$.
The MH algorithm consists of the following steps. 
First starting from $x$, 
we propose a candidate $x^{\prime}$ which is drawn from a certain probability distribution $g(x^{\prime}|x)$
which we call proposal density.    
Then we accept the candidate $x^{\prime}$ with a probability $P_{MH}(x,x^{\prime})$ 
as the next value of the Markov chain:
\be
P_{MH}(x,x^{\prime})= \min\left[1,\frac{P(x^{\prime})}{P(x)}\frac{g(x|x^\prime )}{g(x^{\prime}|x)}\right].
\label{eq:MH}
\ee
If $x^{\prime}$ is rejected we keep the previous value $x$.
Then we repeat these steps.

%When $g(x|x^\prime )=g(x^{\prime}|x)$, eq.(\ref{eq:MH} reduces to the Metropolis accept probability:
%\be
%P_{Metro}(x,x^{\prime})= \min\left[1,\frac{P(x^{\prime})}{P(x)}\right].
%\ee

\section{Adaptive construction scheme}
By choosing an adequate proposal density for the MH algorithm
one may reduce the correlation between the sampled data.  
The posterior density of GARCH parameters often resembles to a Gaussian-like shape. 
Thus one may choose a density similar to a Gaussian distribution as the proposal density.
Such attempts have been done by Mitsui, Watanabe\cite{WATANABE} and Asai\cite{ASAI}.
They used a multivariate Student's t-distribution in order to cover the tails of the posterior density and 
determined the parameters to specify the distribution by 
using the ML technique. 
Here we also use a multivariate Student's t-distribution 
but determine the parameters through MCMC simulations.

The ($p$-dimensional) multivariate Student's t-distribution is given by
\bea
g(\theta)& = & \frac{\Gamma((\nu+p)/2)/\Gamma(\nu/2)}{\det \Sigma^{1/2} (\nu\pi)^{p/2}} \nonumber \\
         &   & \times \left[1+\frac{(\theta-M)^t \Sigma^{-1}(\theta-M)}{\nu}\right]^{-(\nu+p)/2},
\label{eq:ST}
\eea
where $\theta$ and $M$ are column vectors,  
\be
\theta=\left[
\begin{array}{c}
\theta_1 \\
\theta_2 \\
\vdots \\
\theta_p
\end{array}
\right],
M=\left[
\begin{array}{c}
M_1 \\
M_2 \\
\vdots \\
M_p
\end{array}
\right],
\ee
and $M_i=E(\theta_i)$.
$\dis \Sigma$ is the covariance matrix defined as
\be
\frac{\nu\Sigma}{\nu-2}=E[(\theta-M)(\theta-M)^t].
\ee
$\nu$ is a parameter to tune the shape of Student's t-distribution. 
When $\nu \rightarrow \infty$ the Student's t-distribution goes to a Gaussian distribution.
At small $\nu$ Student's t-distribution has a fat-tail.
We also define the matrix $V$ by $\dis V=E[(\theta-M)(\theta-M)^t]$ for later use.

There are 4 parameters in the QGARCH model.
Thus $p=4$, $\dis \theta=(\theta_1,\theta_2,\theta_3,\theta_4)=(\alpha,\beta,\omega,\gamma)$ 
and, $\Sigma$ and $V$ are  $4\times4$ matrices.
The unknown parameters in $M$ and $\Sigma$ are determined by using the data obtained from MCMC simulations.
First we make a short run by the Metropolis algorithm and accumulate some data.
Then we estimate $M$ and $\Sigma$. Note that there is no need to estimate $M$ and $\Sigma$ accurately. 
Second we perform a MH simulation with the proposal density of eq.(\ref{eq:ST}) with the estimated $M$ and $\Sigma$.
After accumulating more data, we recalculate $M$ and $\Sigma$, and update $M$ and $\Sigma$ of eq.(\ref{eq:ST}).
By doing this, we adaptively change the shape of eq.(\ref{eq:ST}) to fit the posterior density.
We call eq.(\ref{eq:ST}) with the estimated $M$ and $\Sigma$ "adaptive proposal density".

\section{Numerical results}
We examine the adaptive construction scheme
by using artificial QGARCH data generated with known parameters.
The QGARCH parameters are set to $\alpha=0.07$, $\beta=0.8$, $\gamma=-0.05$ and $\omega=0.1$.
We have generated 2000 data by the QGARCH process with these parameters as displayed in Fig.1.

\begin{figure}
%\vspace{5mm}
\centering
\includegraphics[height=5cm]{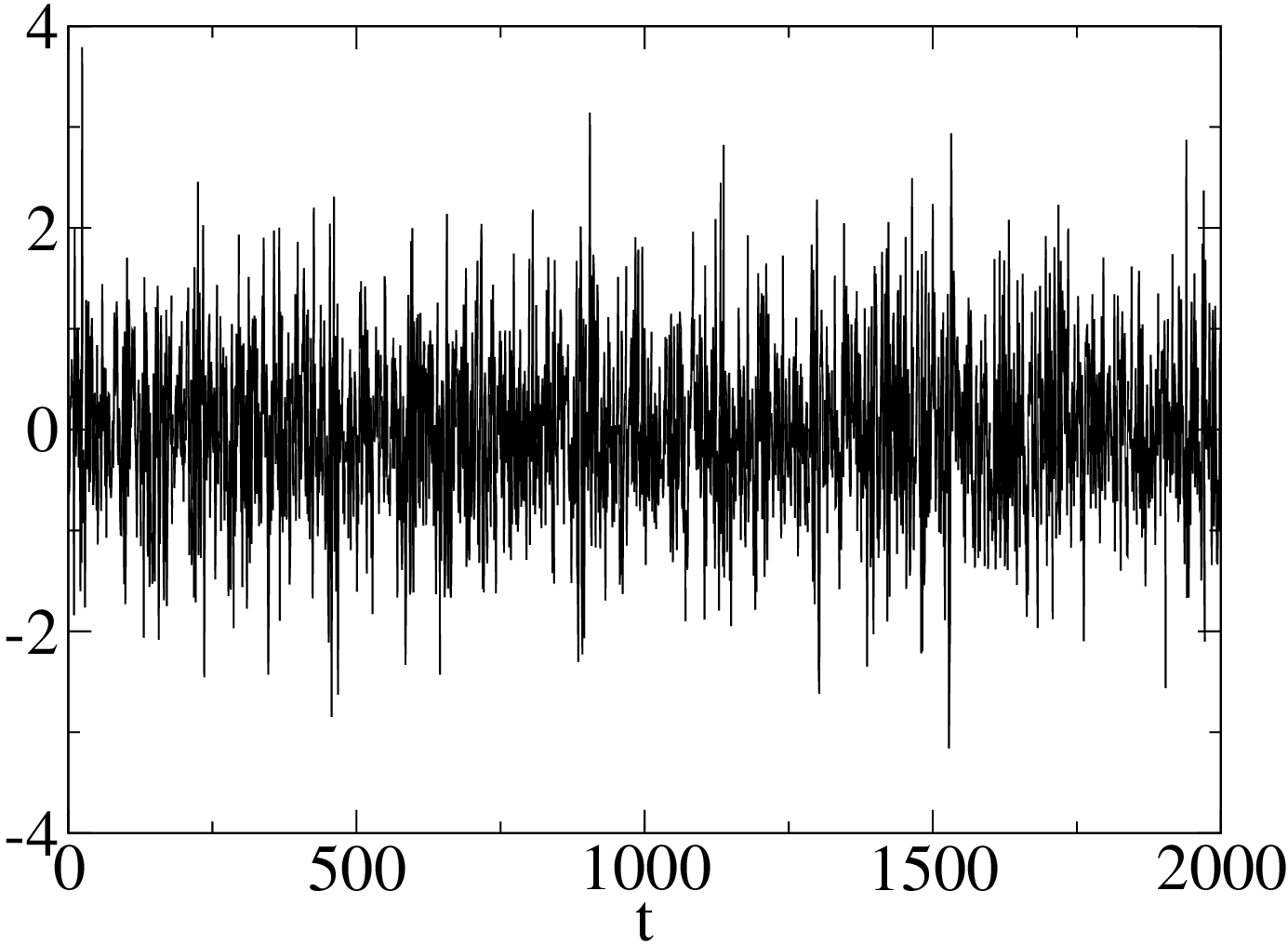}
\caption{
The times series generated by the QGARCH process with  $\alpha=0.07$, $\beta=0.8$, $\gamma=-0.05$ and $\omega=0.1$.
}
%\vspace{1mm}
\label{fig:DATA}
\end{figure}

The adaptive construction scheme is implemented as follows. 
First we start a run by the Metropolis algorithm.
The first 3000 data are discarded as burn-in process or in other words thermalization.
Then we accumulate 1000 data for $M$ and $\Sigma$ estimations.
The estimated $M$ and $\Sigma$ are substituted to $g(\theta)$.
In this study we take $\nu=10$.
We re-start a run by the MH algorithm with $g(\theta)$. 
Every 1000 updates we re-calculate $M$ and $\Sigma$ and update $g(\theta)$.
We accumulate 100000 data for analysis.

\begin{figure}
\vspace{5mm}
\centering
\includegraphics[height=5cm]{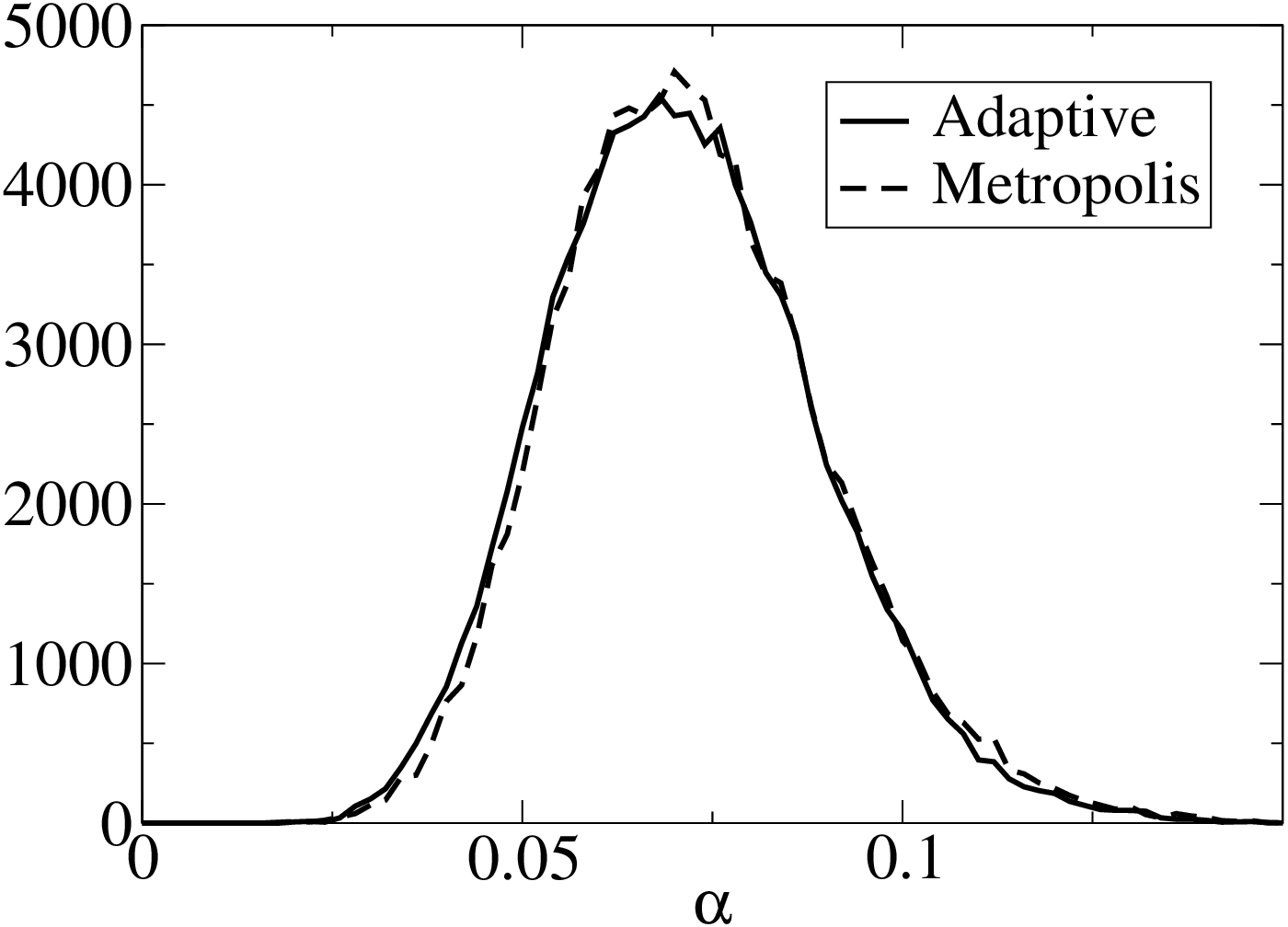}
\caption{
Histograms of $\alpha$ from the adaptive construction scheme
and the Metropolis algorithm.
}
\vspace{1mm}
\label{fig:Histogram}
\end{figure}

For comparison we also use a standard Metropolis algorithm 
to infer the GARCH parameters 
and accumulate 100000 data for analysis.
%In this study the Metropolis algorithm is implemented as follows.
%We draw a candidate $\theta^\prime$ by 
%adding a small random value $\delta \theta$ to the present value $\theta$:
%\be
%\theta^\prime = \theta + \delta \theta,
%\ee
%where $\dis \delta \theta= d(r-0.5)$.
%$r$ is a uniform random number in $[0,1]$ and 
%$d$ is a constant to tune the Metropolis acceptance.
%We choose $d$ so that the acceptance becomes greater than $50\%$.

\begin{figure}
%\vspace{5mm}
\centering
\includegraphics[height=5cm]{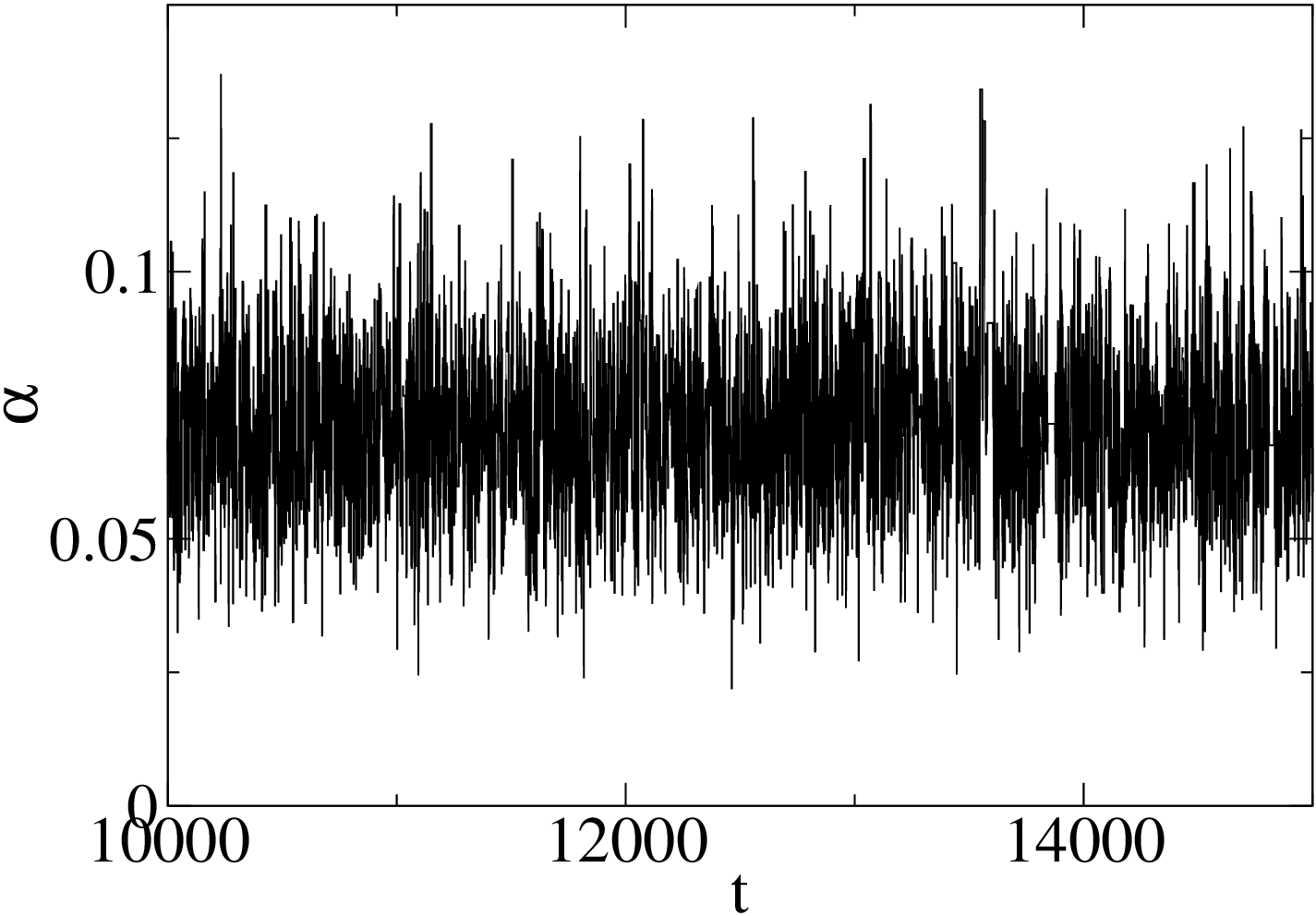}
\caption{
Monte Carlo history of $\alpha$ from the adaptive construction scheme.
}
%\vspace{1mm}
\label{fig:History}
\end{figure}

\begin{figure}
\vspace{5mm}
\centering
\includegraphics[height=5cm]{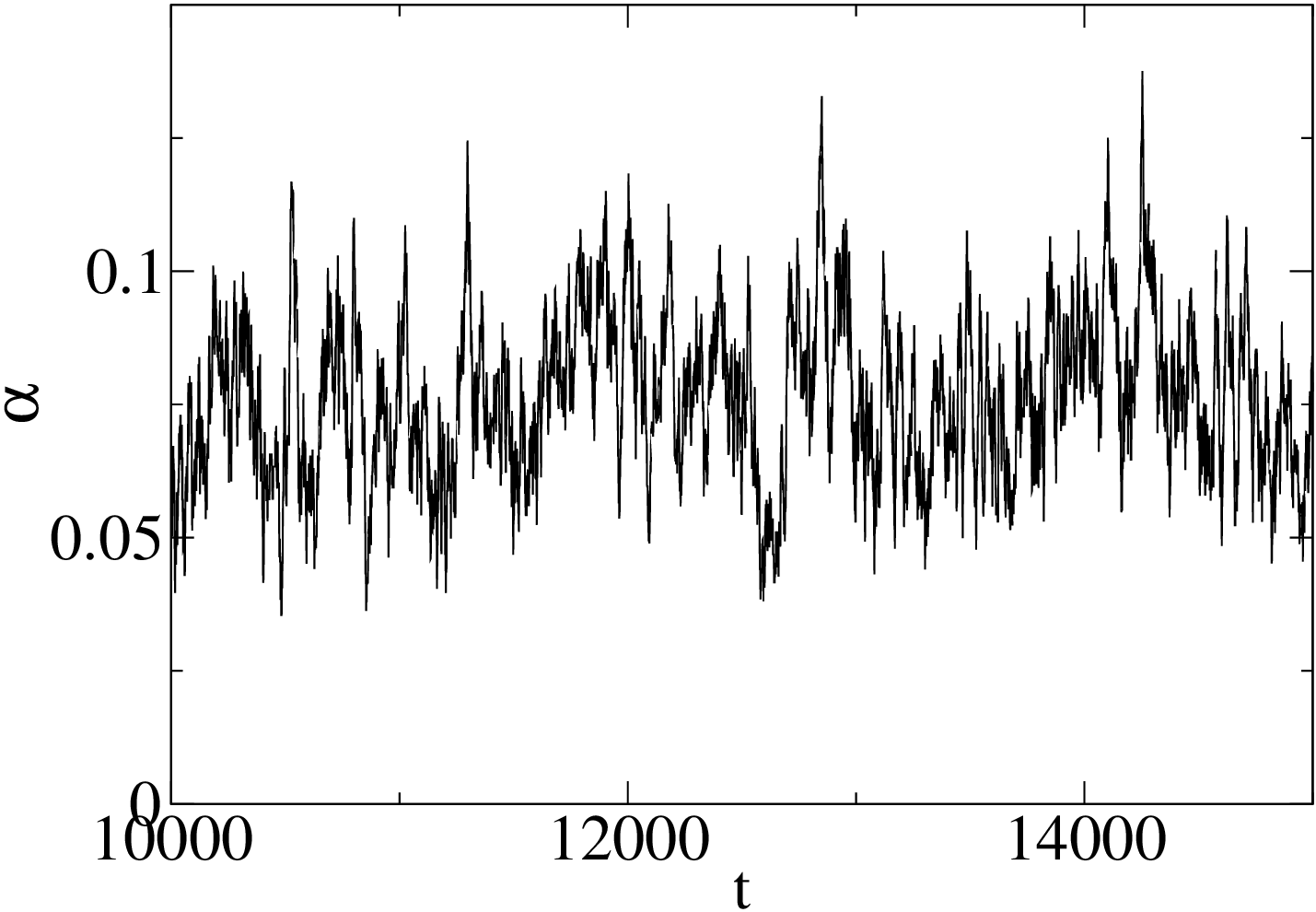}
\caption{
Monte Carlo history of $\alpha$ from the Metropolis algorithm.
}
%\vspace{1mm}
\label{fig:History2}
\end{figure}

\begin{table}[h]
  \centering
  \caption{Results of QGARCH parameters.
   SD and SE stand for standard deviation and statistical error respectively.}
  \label{tab:1}
  {\footnotesize
    \begin{tabular}{cllll}
      \hline
        & \multicolumn{1}{c}{$\alpha$} &
      \multicolumn{1}{c}{$\beta$} &
      \multicolumn{1}{c}{$\omega$}&
      \multicolumn{1}{c}{$\gamma$}         \\
      \hline
\hline
   true & 0.07 & 0.8 & 0.1 & -0.05 \\
\hline
   Adaptive  & 0.07143 & 0.7905 & 0.1054 & -0.04643 \\
   SD        & 0.018   & 0.056  & 0.035  &  0.019   \\
   SE        & 0.00012 & 0.0005 & 0.0004 &  0.00010 \\
   $2\tau$    & $4.1 \pm 1.3$    & $10 \pm 5.0$ & $11\pm 5.1$  & $3.0 \pm 0.4$  \\
\hline
   Metropolis  & 0.0704 & 0.7943 & 0.1032 & -0.0465 \\
   SD          & 0.018  & 0.053  & 0.033  &  0.019  \\
   SE          & 0.0011 & 0.0052 & 0.0032 &  0.0004 \\
   $2\tau$      & $340\pm 100$  & $840\pm 280$  & $820\pm 290$ & $54 \pm7$ \\
      \hline
    \end{tabular}
  }
\end{table}

The results of the parameters inferred by the adaptive construction scheme and 
Metropolis algorithm are summarized in Table 1.
%We see that both methods well reproduce the values of the input parameters.

Fig.~2 compares  the histogram of the sampled data $\alpha$ by the adaptive construction scheme 
with that by the Metropolis algorithm.
The histograms from the two MCMC techniques seem to coincide each others. 
However the property of the sampled data is very different,
which will be analyzed in the followings.

Fig.~3-4 show Monte Carlo time histories from the adaptive construction scheme 
and Metropolis algorithm. 
It is clearly seen that the data sampled by the Metropolis algorithm are substantially correlated. 
The similar behavior is also seen for the sampled data for other parameters. 

The correlations between the data can be measured by the autocorrelation function (ACF). 
The ACF of certain successive data $\theta^{(i)}$ is defined by
\be
ACF(t) = \frac{\frac1N\sum_{j=1}^N(\theta^{(j)}- \bra\theta\ket )(\theta^{(j+t)}-\bra\theta\ket)}{\sigma^2_\theta},
\ee
where $\bra\theta\ket$ and $\sigma^2_\theta$ are the average value and the variance of $\theta$ respectively.

Fig.~5-6 show the ACF of $\alpha$ sampled from the adaptive construction scheme and the Metropolis algorithm.
The ACF of the adaptive construction scheme decreases quickly as Monte Carlo time $t$ increases.
On the other hand the ACF of the Metropolis algorithm decreases very slowly which indicates that
the correlation between the sampled data is very large.

The autocorrelation time (ACT) $\tau$ is calculated by
\be
\tau = \frac12 +\sum_{i=1}^{\infty}ACF(i).
\ee
Results of $\tau$ are summarized in Table 1. 
We find that the ACT from the adaptive construction scheme
have much smaller $\tau$ than those from the Metropolis simulations. 
For instance $\tau$ of $\alpha$ parameter from the adaptive construction scheme is
decreased by a factor of 90 compared to that from the Metropolis algorithm. 
These results  prove that the adaptive construction scheme is 
an efficient algorithm for sampling  de-correlated data.  
The differences in $\tau$ also explain that the statistical errors from the adaptive construction scheme
are much smaller than those from the Metropolis simulations. 

\begin{figure}
\vspace{5mm}
\centering
\includegraphics[height=5cm]{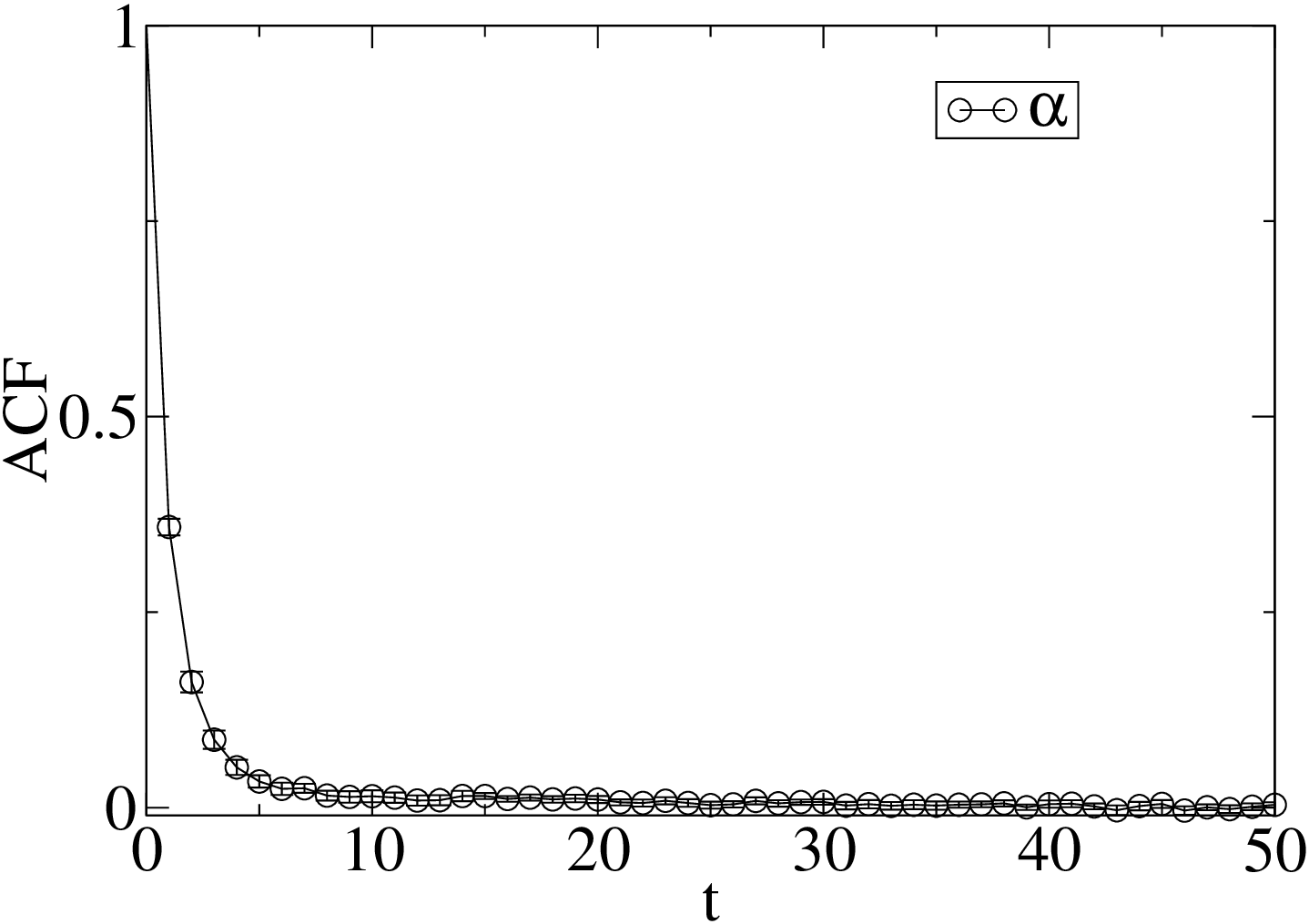}
\caption{
Autocorrelation function of $\alpha$ sampled by the adaptive construction scheme.
}
%\vspace{1mm}
\label{fig:ACF}
\end{figure}

\begin{figure}
\vspace{2mm}
\centering
\includegraphics[height=5cm]{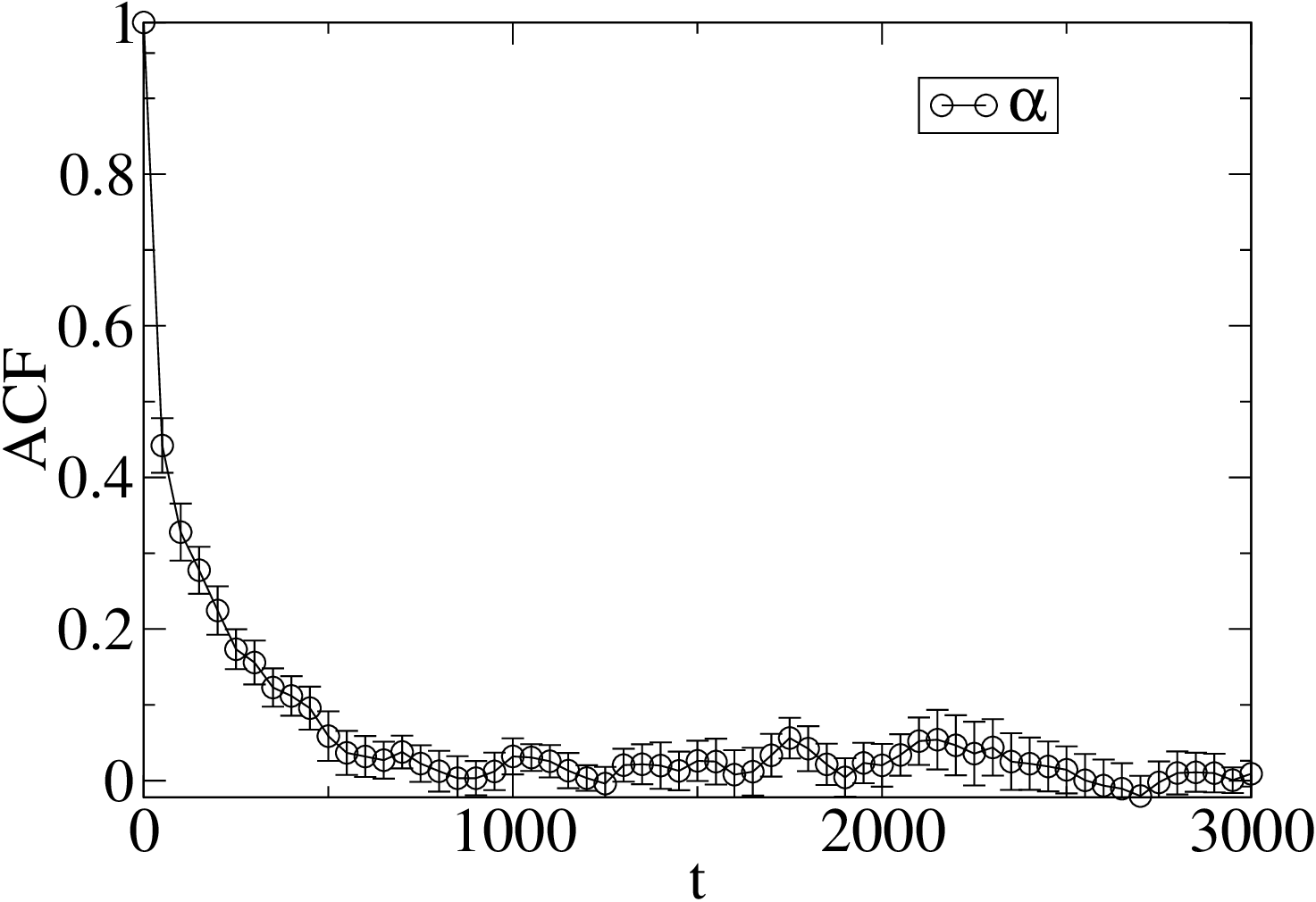}
\caption{
Autocorrelation function of $\alpha$ sampled by the Metropolis algorithm.
}
%\vspace{1mm}
\label{fig:ACF2}
\end{figure}

\begin{figure}[t]
\vspace{5mm}
\centering
\includegraphics[height=5cm]{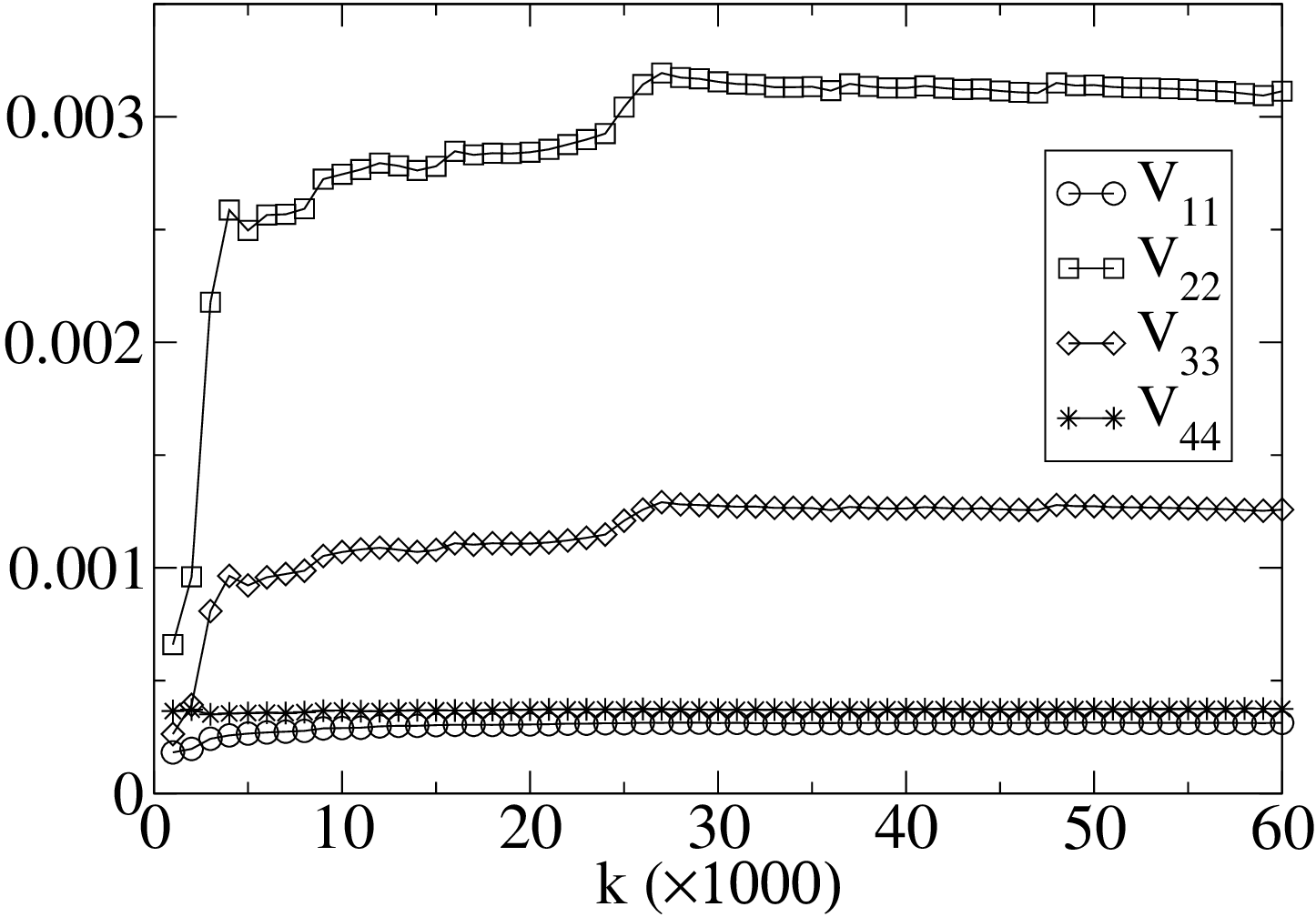}
\caption{
The diagonal elements of $V$ as a function of the data size.
}
\label{fig:SIG}
\end{figure}

\begin{figure}
\vspace{2mm}
\centering
\includegraphics[height=5cm]{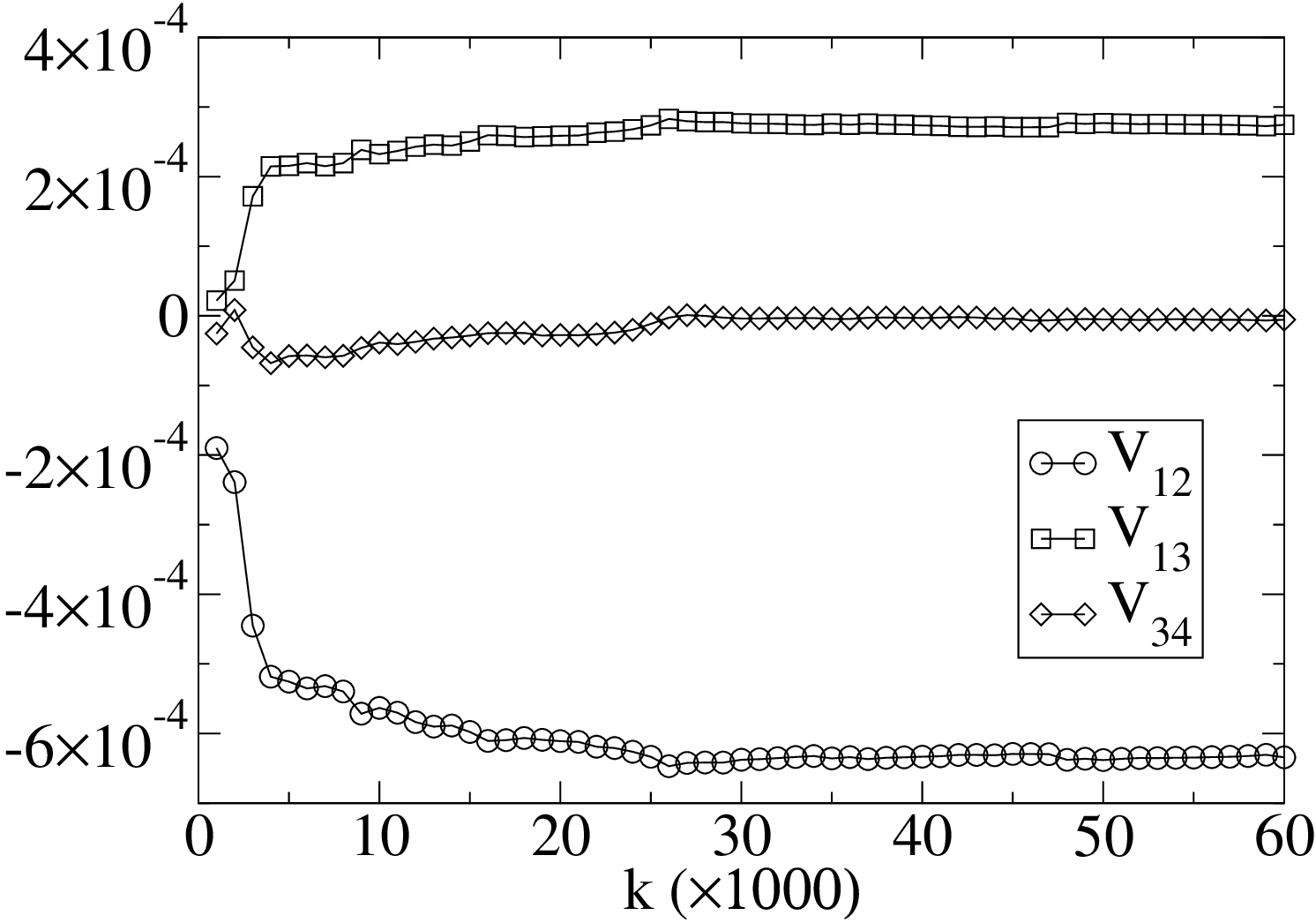}
\caption{
Some off-diagonal elements of $V$ as a function of the data size.
Other off-diagonal elements show the similar convergence behavior.
}
\label{fig:SIG2}
\end{figure}

Fig.~7-8 show the convergence property of the covariance matrix $V$.
%Here let us define a symmetric matrix $V$ as $\dis V=E[(\theta-M)(\theta-M)^t]$ 
%and $\theta=(\theta_1,\theta_2,\theta_3,\theta_4)=(\alpha,\beta,\omega,\gamma)$.
All the elements of $V$ seem to converge to certain values as 
the simulations are proceeded.

\begin{figure}
\vspace{5mm}
\centering
\includegraphics[height=5.0cm]{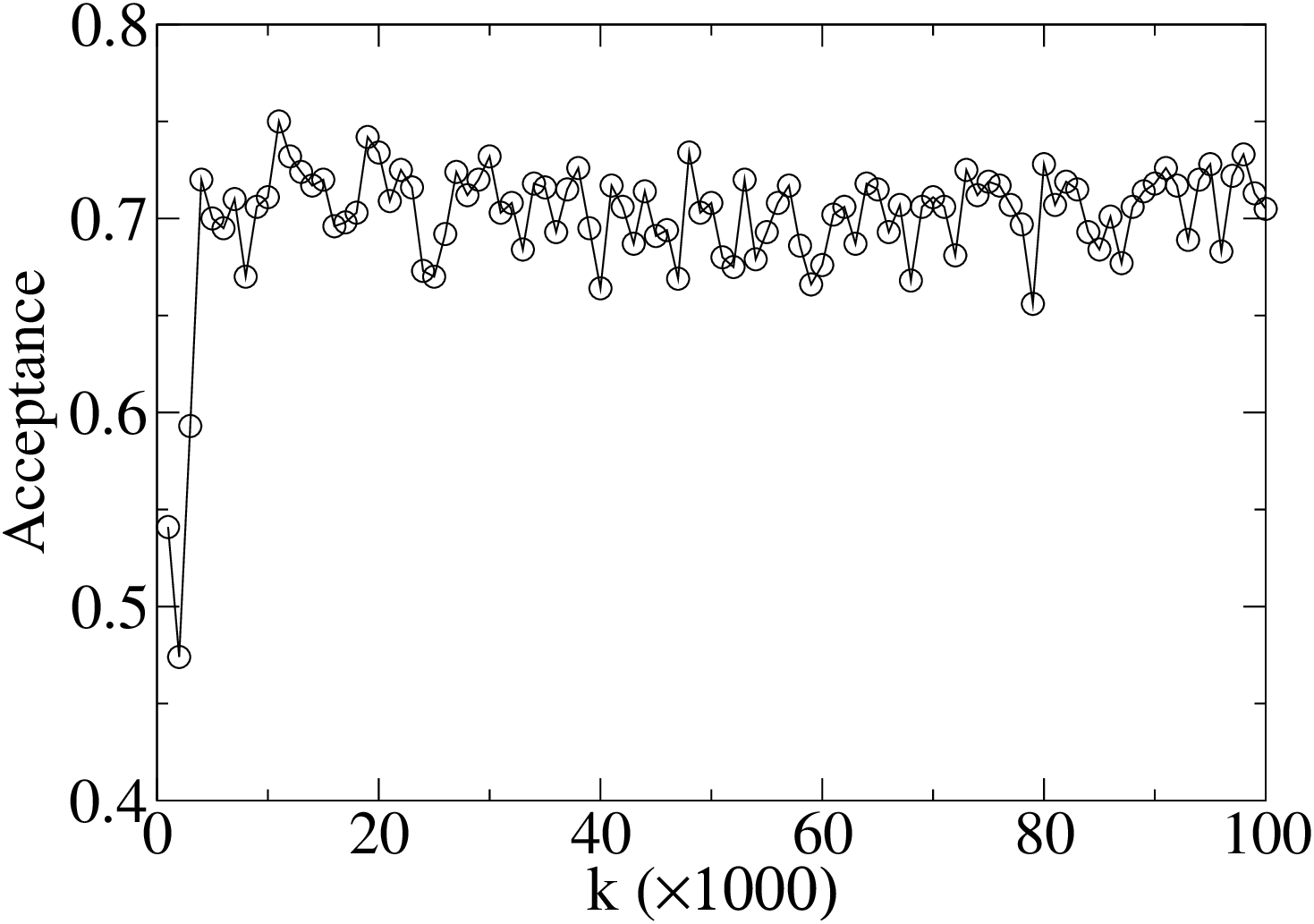}
\caption{
Acceptance at MH step with the adaptive proposal density.
}
\vspace{1mm}
\label{fig:ACC}
\end{figure}

Fig.~9  shows values of the acceptance at the MH algorithm with the adaptive proposal density of eq.(\ref{eq:ST}).
Each acceptance is calculated every 1000 updates and the calculation of the acceptance is
based on the latest 1000 data.
At the first stage of the simulation
the acceptance is low. This is  because $M$ and $\Sigma$ have not yet been calculated accurately as we see in Fig.~7-8.
However the acceptances increase quickly as the simulations are proceeded
and reaches plateaus of about $70\%$. This acceptance is reasonably high for the MH algorithm.

%Since the acceptance quickly reaches a plateau of about $70\%$, we may not continue to update 
%the paremeters of the Student's t-distribution during the simulation.    

\section{Conclusions}
We have performed the Bayesian inference of the QGARCH model by 
applying the adaptive construction scheme to the MH algorithm. 
The adaptive construction scheme, which does not use the ML method, is found to be very efficient in the sense that
the autocorrelation times of the sampled data are very small. 
Thus the adaptive construction scheme serves as an efficient MCMC technique for 
the Bayesian inference of the QGARCH model.
The adaptive construction scheme is not limited to the Bayesian inference of the QGARCH or GARCH models  
and can be applied for other GARCH-type models.

Although we have updated the parameters of the Student's t-distribution
adaptively using the sampled data during the simlation, 
we may take a strategy that we stop to update the parameters at some point.
As seen in fig. 9, the acceptance quickly reaches a plateau of about $70\%$ 
and after that the adaptive parameter update does not improve the acceptance.
Thus we may stop to update the parameters and use 
the fixed proposal density for the MH step after the acceptance reaches the plateau.

\section*{Acknowledgments}
The numerical calculations were carried out on Altix at the Institute of Statistical Mathematics
and on SX8 at the Yukawa Institute for Theoretical Physics 
in Kyoto University. 
%The autor thank Omori for comments on the recent progress on the adaptive Monte Carlo methods.
%This study was carry out under the ISM Cooperative Research Program (

\end{document}